\documentclass[a4paper]{article}
\usepackage{amsmath,amssymb,graphicx}

\newcommand{\dt}{\! \cdot \!}

\newcommand{\FUN}[1]{{\mathrm{#1}}}
\newcommand{\EMB}{\boldsymbol}
\newcommand{\VEC}[1]{{\EMB{#1}}}
\newcommand{\ALG}[1]{{\mathcal{#1}}}
\newcommand{\LAB}[1]{{\mathsf{#1}}}
\newcommand{\SIG}[1][{}]{\EMB\sigma_{\!#1}}
\newcommand{\ISIG}[1][{}]{\EMB\iota\!\EMB\sigma_{\!#1}}
\newcommand{\KET}[1]{|\,#1\,\rangle}
\newcommand{\BRA}[1]{\langle\,#1\,|}

\newcommand{\HALF}{\mathchoice{\textstyle\frac12}{\frac12}{\frac12}{\frac12}}
\newcommand{\RHO}{{\EMB\rho}}

\newcommand{\OL}{\overline}

\newcommand{\IMJ}{\EMB j}

\begin{document}

\begin{center}

{\bf\Large Interaction and Entanglement in the \\ Multiparticle Spacetime
Algebra}

\vspace{0.4cm}

Timothy F.~Havel$^1$ and Chris J.~L.~Doran$^2$

\vspace{0.4cm}

{}$^1$MIT (NW14-2218), 150 Albany St., Cambridge, MA 02139, USA. 
\texttt{tfhavel@mit.edu}

{}$^2$Astrophysics Group, Cavendish Laboratory, Madingley Road, 
Cambridge CB3 0HE, UK. \texttt{c.doran@mrao.cam.ac.uk}  

\vspace{0.4cm}

\begin{abstract}
The multiparticle spacetime algebra (MSTA) is an extension of Dirac
theory to a multiparticle setting, which was first studied by Doran,
Gull and Lasenby.  The geometric interpretation of this algebra, which
it inherits from its one-particle factors, possesses a number of
physically compelling features, including simple derivations of the
Pauli exclusion principle and other nonlocal effects in quantum
physics.  Of particular importance here is the fact that all the
operations needed in the quantum (statistical) mechanics of spin $1/2$
particles can be carried out in the ``even subalgebra'' of the MSTA.
This enables us to ``lift'' existing results in quantum information
theory regarding entanglement, decoherence and the quantum / classical
transition to space-time.  The full power of the MSTA and its
geometric interpretation can then be used to obtain new insights into
these foundational issues in quantum theory.  A system of spin $1/2$
particles located at fixed positions in space, and interacting with an
external magnetic field and/or with one another via their intrinsic
magnetic dipoles provides a simple paradigm for the study of these
issues.  This paradigm can further be easily realized and studied in
the laboratory by nuclear magnetic resonance spectroscopy.
\end{abstract}

\end{center}

\section{The Physics of Quantum Information}

\emph{Information}, to be useful, must be encoded in the state of a
physical system.  Thus, although information can have more than one
meaning, it always has something to say about the state of the system
it is encoded in.  Conversely, the average information needed to
specify the state of a system drawn at random from some known
probability distribution is proportional to the \emph{entropy} of the
corresponding statistical mechanical ``system''.  It follows that
entropy can be understood as a measure of the system's information
storage capacity.  The physics of information is a fertile area of
research which promises to become increasingly important as computers
and nanotechnology approach the limits of what is physically possible
\cite{BenneLanda:85,FeyAllHey:00}.

In practice, information is usually binary encoded into an array of
two-state systems, each of which can hold one bit of data, where the
two ``states'' correspond to the minimum or maximum value of a
continuous degree of freedom.  Ordinarily such physical \emph{bits}
obey classical mechanics, but many examples of two-state quantum
systems are known, for example adjacent pairs of energy levels in
atoms, photon polarizations, or the magnetic dipole orientations of
fermions.  These quantum bits, or \emph{qubits} as they are called,
have a number of distinctive and nonintuitive properties
\cite{BenneDivin:00,NielsChuan:00,WilliClear:99}.  In particular, the
number of parameters needed to specify (the statistics of measurement
outcomes on) the \emph{joint} state of an array of $n$ qubits is
generally $2^n$ --- exponentially larger than the $n$ needed for an
array of classical bits!  These new degrees of freedom are due to the
existence of non-separable or \emph{entangled} states, which may be
thought of as providing direct paths between pairs of states related
by flipping more than one bit at a time.  A further mysterious
property of these quantum states stems from the fact that it is
impossible (so far as is known!) to determine just where on the
pathways between states the qubits are.  This is because the act of
\emph{measuring} the qubits' states (however this may be done) always
``collapses'' them into one of their extremal states \cite{Peres:93}.

A \emph{quantum computer} is an array of distinguishable qubits that
can be put into a known state, evolved under a sequence of precisely
controlled interactions, and then measured.  It has been shown that
such a computer, \emph{if} one could be built, would be able to solve
certain problems asymptotically more rapidly than any classical
device.  Unfortunately, quantum systems are exceedingly difficult to
isolate, control and monitor, so that at this time only simple
prototype quantum computers have been operated in the laboratory.
Although far from competitive with today's laptops, these prototypes
are of great scientific interest.  This is because quantum computers
provide a paradigm for the study of a number of poorly understood
issues in quantum mechanics, including \emph{why} the particular
classical world we inhabit is singled out from the myriads allowed by
quantum mechanics.  This is widely believed to be the result of
\emph{decoherence}: the decay of accessible information due to the
entanglement generated by the interactions between the system with its
environment \cite{GiuliniEtAl:96,PazZurek:01,Percival:98}. The reason
such issues are poorly understood, even though the microscopic laws of
quantum mechanics are complete and exact, lies in our very limited
ability to integrate these laws into precise solutions for large and
complex quantum systems, or even to gain significant insights into
their long-term statistical behavior.  The intrinsic complexity of
quantum dynamics is in fact precisely what makes quantum computers so
powerful to begin with!

This paper will explore the utility of the \emph{multiparticle
spacetime algebra} (MSTA) description of qubit states, as introduced
by Doran, Gull and Lasenby
\cite{DorLasGul:93,DoLaGuSoCh:96,SomLasDor:99}, for the purposes of
understanding entanglement, decoherence and quantum complexity more
generally.  As always with applications of geometric algebra, our goal
will be to discover simple geometric interpretations for otherwise
incomprehensible algebraic facts.  To keep our study concrete and our
observations immediately amenable to experimental verification, we
shall limit ourselves to the qubit interactions most often encountered
in physical implementations, namely the interaction between the
magnetic dipoles of spin $1/2$ particles such as electrons, neutrons
and certain atomic nuclei.  These will be assumed to have fixed
positions, so their spatial degrees of freedom can be ignored.
Examples of such systems, often involving $10^{20}$ or more spins
(qubits), are widely encountered in chemistry and condensed matter
physics, and can readily be studied by various spectroscopies,
most notably nuclear magnetic resonance (NMR)
\cite{Abragam:61,HaCoSoTs:00,Slichter:90,Weiss:99}.

\section{The Multiparticle Space-Time Algebra}

It will be assumed in the following that the reader has at least a
basic familiarity with geometric algebra, as found in e.g.\
\cite{Baylis:96,DorLasGul:93,HestNF1:99}.  This brief account of the
MSTA is intended mainly to introduce the notation of the paper, while
at the same time providing a taste of how the MSTA applies to quantum
information processing.  More introductory accounts may be found in
\cite{DoLaGuSoCh:96,HaCoSoTs:00,HavelDoran:01,SomCorHav:98}.

The $n$-particle MSTA is the geometric algebra $\ALG G(n,3n)$
generated by $n$ copies of Minkowski space-time $\ALG R^{1,3}$.
We let $\{\gamma_\mu^a\,|\,\mu=0,\ldots,3\}$ denote a basis set
of vector generators for this algebra, satisfying
\begin{equation}
\gamma_\mu^a \dt \gamma_\nu^b ~=~ \eta_{\mu\nu\,} \delta^{ab} \,.
\end{equation}
The superscripts refer to separate particle spaces ($a,b=1,\ldots,n$)
and the subscripts label spacetime vectors ($\mu,\nu=0,\ldots,3$), while
$\eta_{\mu\nu}$ is the standard Minkowski metric of signature $(+--\,-)$.
The $\{\gamma_\mu^a\}$ can be thought of as a basis for relativistic
\emph{configuration space}, and the MSTA is the geometric algebra of
this space.

The even subalgebra of each copy of the spacetime algebra is
isomorphic to the algebra of Euclidean three-dimensional space $\ALG
G(3)$ \cite{Hestenes:66}.  The specific map depends on a choice
of timelike vector, and the algebra then describes the rest space
defined by that vector.  It is convenient in most applications to
identify this vector with $\gamma_0$, and we define
\begin{equation}
\SIG[k]^a ~=~ \gamma_k^a \gamma_0^a \,.
\end{equation}
Each set of \emph{spatial vectors} $\SIG[1]^a, \SIG[2]^a, \SIG[3]^a$
generates a three-dimensional geometric algebra.  It is easily seen
that the generators of different particle spaces \textit{commute},
so that the algebra they generate is isomorphic to the Kronecker product
$\ALG G(3)\,\otimes\cdots\otimes\,\ALG G(3)$.  For this paper we will work
almost entirely within this (non-relativistic) space, but it should be
borne in mind throughout that all results naturally sit in a fully
relativistic framework.

To complete our definitions, we denote the pseudoscalar for each particle by
\begin{equation}
{\EMB \iota}^a ~=~ \gamma_0^a \gamma_1^a \gamma_2^a \gamma_3^a ~=~
\SIG[1]^a \SIG[2]^a \SIG[3]^a \,.
\end{equation}
For the bivectors in each spatial algebra we make the abbreviation
${\EMB \iota}^a \SIG[k]^a = \ISIG[k]^a$.  The reverse operation in the
MSTA is denoted with a tilde.  This flips the sign of both vectors and
bivectors in $\ALG G(3)$, and so does not correspond to spatial reversion.

There are many ways to represent multiparticle states within the MSTA.
Here we are interested in an approach which directly generalizes
that of Hestenes for single-particle non-relativistic states.  We
will represent states using multivectors constructed from products of
the even subalgebras of each $\ALG G(3)$.  That is, states are
constructed from sums and products of the set $\{1, \ISIG[k]^a\}$,
where $k=1\ldots 3$ and $a$ runs over all particle spaces.  This
algebra has real dimension $4^n$, which is reduced to the expected
$2^{n+1}$ by enforcing a consistent representation for the unit
imaginary.  This is ensured by right-multiplying all states with the
\emph{correlator} idempotent
\begin{equation}
\VEC E ~\equiv~
\HALF(1 - \ISIG[3]^1\ISIG[3]^2)\,
\HALF(1 - \ISIG[3]^1\ISIG[3]^3)
~\cdots~
\HALF(1 - \ISIG[3]^1\ISIG[3]^n) ~.
\end{equation}
The correlator is said to be \textit{idempotent} because it satisfies
the projection relation
\begin{equation}
\VEC{E}^2 ~=~ \VEC{E}.
\end{equation}
In the case of two spins, for example, we have $\VEC E =
-\ISIG[3]^1\ISIG[3]^2\VEC E$, so that any term right-multiplied by
$-\ISIG[3]^1\ISIG[3]^2$ is projected by $\VEC E$ back to the same
element.  The correspondence with the usual complex vector space
representation is obtained by observing that (for e.g.~two spins)
every spinor may be written uniquely as
\begin{equation}
\EMB\psi ~=~ \left( \EMB\psi_1 - \ISIG[2]^2 \EMB\psi_2 - \ISIG[2]^1
\EMB\psi_3 + \ISIG[2]^1\ISIG[2]^2 \EMB\psi_4 \right) \VEC E ~,
\end{equation}
where the $\EMB\psi_k$ are ``complex numbers'' of the form $\alpha_k +
\beta_k \VEC J$ with $\alpha_k$, $\beta_k$  real, and the role of the
imaginary is played by the (non-simple) bivector
\begin{equation}
\VEC J ~\equiv~ \ISIG[3]^1 \VEC E ~=~
\ISIG[3]^2 \VEC E ~=~ \HALF (\ISIG[3]^1 + \ISIG[3]^2).
\end{equation}
The complex generator $\VEC{J}$ satisfies $\VEC{J}^2 ~=~ -\VEC{E}$,
which ensures consistency with the standard formulation of quantum
theory.  As in the single-particle case, the complex structure is
always represented by \textit{right-multiplication} by $\VEC{J}$.
While this approach may look strange at first, it does provide a
number of new geometric insights into the nature of multiparticle
Hilbert space \cite{DoLaGuSoCh:96}.

\section{Two Interacting Qubits}

\begin{figure}
\begin{minipage}{0.45\textwidth}
\includegraphics[scale=0.2]{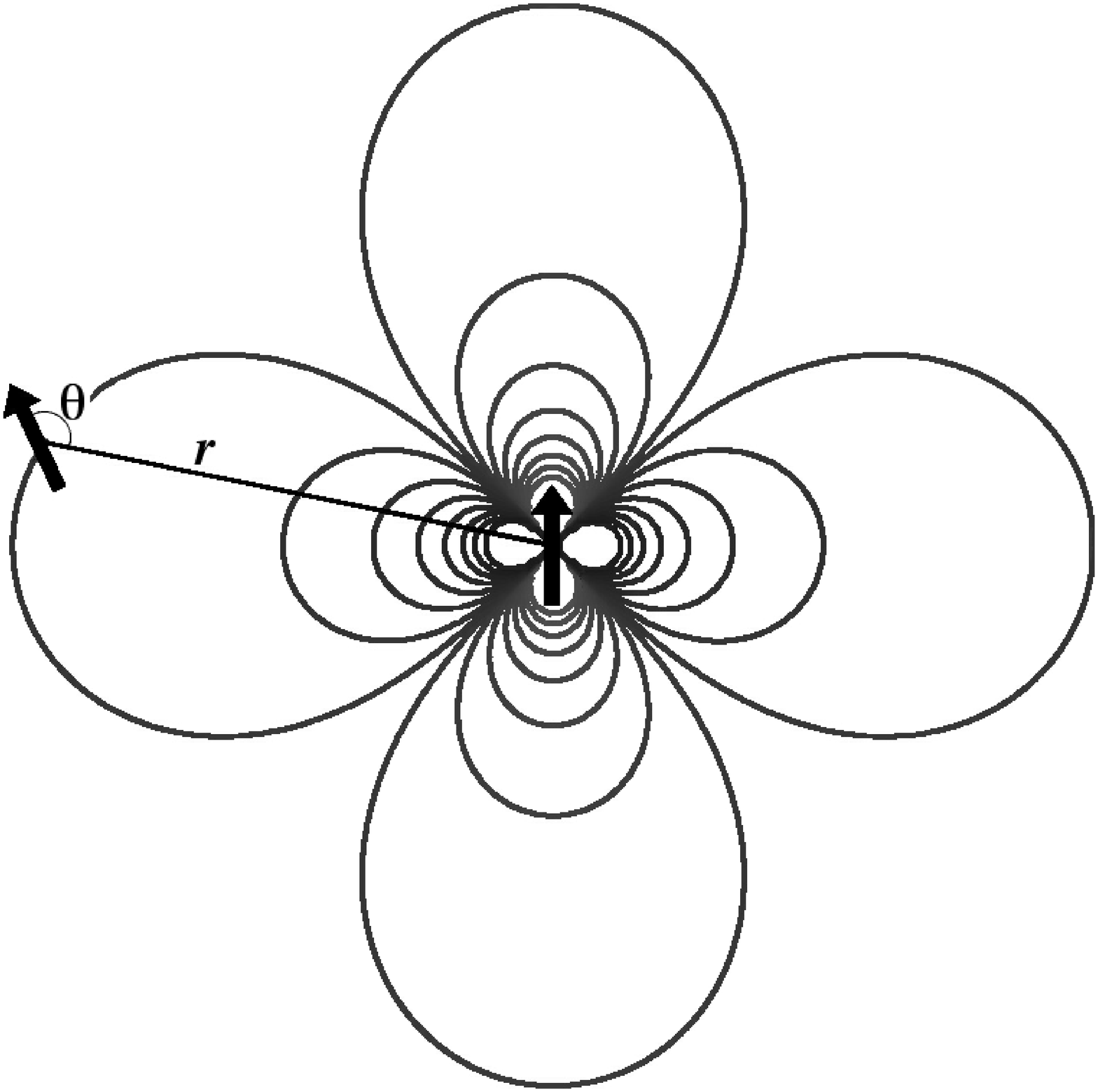}
\end{minipage} \hspace{0.05\textwidth}
\begin{minipage}{0.45\textwidth}
\includegraphics[scale=0.40]{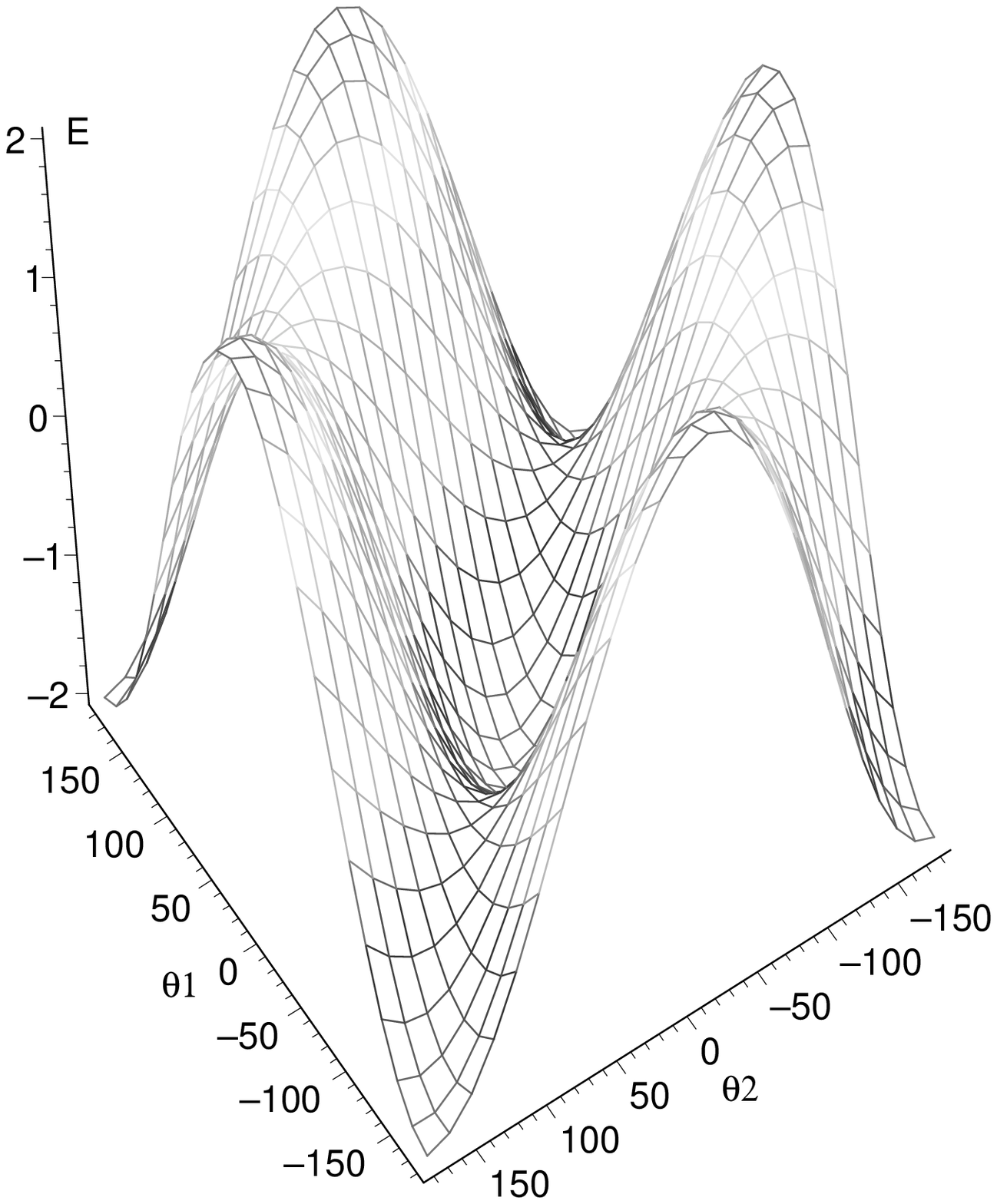}
\end{minipage}
\caption{One dipole in the field of another,
the radial vector $\VEC r$ between them, and the
angle $\theta$ it makes with the dipole (left);
corresponding potential energy surface
(assuming the dipoles are confined
to the plane as depicted on the left)
showing that there are two minima, two
maxima, and four saddle points (right).}
\label{fig:dd}
\end{figure}

As an application of the MSTA approach, we consider a simple model
system of interacting qubits.  This exhibits all of the complexity
of multiparticle quantum systems generally, including the role of
entanglement and the distinction between classical and quantum theories.
Associated with the intrinsic angular momentum of any spin $1/2$
particle is a magnetic dipole $\EMB\mu$.  In the far-field limit,
the energy for the interaction between two such dipoles is given
classically by the expression
\begin{equation}
E_D ~=~ \frac{\mu_0}{4\pi} \left(
\frac{\EMB\mu_1 \dt \EMB\mu_2}{r^3} ~-~ 3\,
\frac{\EMB\mu_1 \dt \VEC r \, \EMB\mu_2 \dt \VEC r}{r^5}
 \right) ~,
\label{ClsE}
\end{equation}
where $\mu_0$ is the permeability of the vacuum, $\VEC r$ is the radial
vector between the dipoles, and $r=|\VEC r |$ (see Fig.\ \ref{fig:dd}).
To obtain the quantum theory of this system (via ``first quantization'')
we replace the magnetic moment $\EMB\mu$ by its operator equivalent,
given by the component relation
\begin{equation}
\hat{\mu}_k ~=~ \gamma \hat{s}_k.
\end{equation}
Here $\gamma$ is the gyromagnetic ratio and $\hat{s}_k$ is the spin
operator in the $k$-th direction.  For a spin $1/2$ particle the spin
operators are simply
\begin{equation}
\hat{s}_k ~=~ \HALF \hbar \hat{\sigma}_k,
\end{equation}
where the $\hat{\sigma}_k$ are the Pauli matrix operators.  The
classical energy of~(\ref{ClsE}) gives rise to a quantum Hamiltonian
containing 2 terms.  The first involves $\EMB\mu_1 \dt \EMB\mu_{2\,}$,
which is replaced by the operator 
\begin{equation}
\sum_{k=1}^3 \hat{\mu}_k^1 \, \hat{\mu}_k^2 ~=~
\frac{\hbar^2\gamma_1\gamma_2}{4} \sum_{k=1}^3 \, \hat{\sigma}_k
\otimes \hat{\sigma}_k,
\end{equation}
where $\gamma_1$ and $\gamma_2$ are the gyromagnetic ratios of
particles 1 and 2 respectively.  This operator acts on a four-dimensional
complex vector space.  To form the MSTA equivalent of this we write
\begin{equation}
\sum_{k=1}^3 \, \hat{\sigma}_k \otimes \hat{\sigma}_k
~=~ - \sum_{k=1}^3 \, i\hat{\sigma}_k \otimes i\hat{\sigma}_k ~,
\end{equation}
where $i$ is an (uninterpreted) imaginary unit.
Each factor of $i\hat{\sigma}_k$ has an equivalent action
in the MSTA given by left-multiplication with $\ISIG[k]$
in the corresponding particle's space. 
It follows that we can replace
\begin{equation}
\sum_{k=1}^3 \, \hat{\sigma}_k \otimes \hat{\sigma}_k \KET{\psi}
~\mapsto~ - \sum_{k=1}^3 \ISIG[k]^1 \, \ISIG[k]^2 \, \EMB\psi ~.
\end{equation}
We can already see, therefore, that the Hamiltonian
is going to become a 4-vector in the MSTA.

For the second term we let $\VEC n = \VEC r /r$, so that $\VEC n$ is
the unit vector parallel to the line between the dipoles' centers.
We next form the operator for
$\EMB\mu_1 \!\cdot \VEC n \, \EMB\mu_2 \!\cdot \VEC n$, which is
\begin{equation}
\sum_{k,\ell=1}^3  n_k n_\ell \, \hat{\mu}_k^1 \, \hat{\mu}_\ell^2
~=~ \frac{\hbar^2\gamma_1\gamma_2}{4} \sum_{k,\ell=1}^3  n_k \, n_\ell
\, \hat{\sigma}_k \otimes \hat{\sigma}_\ell \,.
\end{equation}
The MSTA equivalent of this is simply
\begin{equation}
- \frac{\hbar^2\gamma_1\gamma_2}{4} \sum_{k,\ell=1}^3  n_k \, n_\ell \,
\ISIG[k]^1 \, \ISIG[\ell]^2 ~=~ - \frac{\hbar^2\gamma_1\gamma_2}{4} \,
\EMB{\iota}\! {\VEC n}^1 \, \EMB{\iota} \! {\VEC n}^2 \,.
\end{equation}
The role of the Hamiltonian in the MSTA is therefore assumed by the
4-vector
\begin{equation}
\VEC H_\LAB{D} ~\equiv~
- \frac{d}{4} \, \Bigl( \sum_{k=1}^3\, \ISIG[k]^1 \ISIG[k]^2 \,-\,
3\, \EMB{\iota}\! {\VEC n}^1 \EMB{\iota}\! {\VEC n}^2 \Bigr) \,,
\end{equation}
where $d \equiv \mu_0\hbar\gamma_1\gamma_2/4\pi r^3$ $sec^{-1}$.
This definition of $\VEC H_\LAB{D}$ is chosen so that Schr\"odinger's
equation takes the simple form
\begin{equation}
-\partial_t \, \EMB\psi ~=~  \VEC H_\LAB{D} \, \EMB\psi \, \VEC J \,.
\label{SchE}
\end{equation}
>From this one can immediately see a key feature of the MSTA
approach, which is that all references to the tensor product
have been removed.  All one ever needs is the geometric product,
which inherits all of the required properties from the relativistic
definition of the MSTA.

\subsection{The Propagator}

In the more conventional, matrix-based approach
to quantum theory the propagator in the current
setup would be simply $\exp(-i\hat{H}t/\hbar)$.
In finding the MSTA equivalent of this we appear to have a
serious problem.  In equation~(\ref{SchE}) the 4-vector $\VEC
H_\LAB{D}$ acts from the left on $\EMB \psi$, whereas $\EMB J$ sits
on the right.  In fact, this is more of a notational issue than a
foundational one.  We simply define an operator $\IMJ$ to denote
right-multiplication by $\EMB J$ when acting on a multiparticle state,
namely
\begin{equation}
\IMJ \EMB \psi ~\equiv~ \EMB \psi \EMB J \,.
\end{equation}
We are then free to write $\IMJ$ anywhere we chose within a multiplicative
term, and to let it distribute over addition like a multiplicative operator.
This notational device is extremely useful in practice, though occasionally
one has to be careful in applying it.

The propagator can be obtained in a number of ways.  One is to write
\begin{equation}
\VEC H_\LAB{D} ~=~ \frac{d}{2}\, \EMB\Pi \,-\, \frac{d}{4} \,+\,
\frac{3d}{4}\, \EMB{\iota}\! {\VEC n}^1 \EMB{\iota}\! {\VEC n}^2 \,,
\end{equation}
where 
\begin{equation}
\EMB\Pi ~=~ \HALF\bigl( 1 -  \ISIG[1]^1\ISIG[1]^2 -
\ISIG[2]^1\ISIG[2]^2 - \ISIG[3]^1\ISIG[3]^2 \bigr) \,.
\end{equation}
The multivector $\EMB\Pi$ constitutes a geometric
representation of the particle interchange operator,
since $\EMB\Pi\,\EMB\zeta^1\EMB\upsilon^2 = \EMB\upsilon^1
\EMB\zeta^2\EMB\Pi$, and hence satisfies $\EMB\Pi^2=1$. 
It follows that $\EMB\Pi$ commutes with
$\EMB{\iota}\!{\VEC n}^1 \EMB{\iota}\!{\VEC n}^2$,
and the propagator can be written as
\begin{equation} 
\EMB\exp(- \IMJ\, \VEC H_\LAB{D\,} t) ~=~
\EMB\exp(-\IMJ\, dt/4) \, \EMB\exp( \IMJ\, \EMB\Pi\, dt/2 )\,
\EMB\exp( \IMJ\, \EMB{\iota}\! {\VEC n}^1 \EMB{\iota}\!
{\VEC n}^{2\,} 3dt/4  ) .
\label{Prop1}
\end{equation}
All three exponentials in this expression commute.

Alternatively, one can look for eigenstates of the Hamiltonian.  At
this point it is convenient to choose a coordinate system in which
${\VEC r}$ is parallel to the $\LAB z$-axis, so that
\begin{equation}
\VEC n ~=~ \SIG[3]\,, \quad
\VEC H_\LAB{D} ~=~ \frac{d}4\,
\bigl( 2\,\ISIG[3]^1\ISIG[3]^2 \,-\,
\ISIG[1]^1\ISIG[1]^2 \,-\, \ISIG[2]^1\ISIG[2]^2
\bigr) ~.
\end{equation}
One can ``diagonalize'' the Hamiltonian operator in this coordinate
frame by defining
\begin{align}
\VEC H_\LAB{D}'
~=~&~ \VEC e^{(\pi/4)\, \IMJ \ISIG[1]^1\ISIG[2]^2\,}
\VEC H_\LAB{D}\, \VEC e^{-(\pi/4)\,\IMJ \ISIG[1]^1\ISIG[2]^2}
\nonumber \\ 
~=~&~ \frac{d}4\, \Big( 2\, \ISIG[3]^1 \ISIG[3]^2 \,-\,
\IMJ \ISIG[1]^1\ISIG[2]^2\, \big(
\ISIG[1]^1\ISIG[1]^2 + \ISIG[2]^1\ISIG[2]^2 \big) \Big)
\nonumber \\ 
~=~&~ \frac{d}4\, \Big( 2\, \ISIG[3]^1 \ISIG[3]^2 \,+\,
\IMJ  \big( \ISIG[3]^1 - \ISIG[3]^2 \big)
\Big) ~.
\end{align}
The eigenvalues and eigenspinors of $\VEC H_\LAB{D}'$ are
\begin{align}
\VEC H_\LAB{D}' \,  \VEC E ~=~& -\frac{d}2\, \VEC E ~, &
\VEC H_\LAB{D}' \, \ISIG[2]^2 \VEC E ~=~&~ d\, \ISIG[2]^2 \VEC E ~,
\nonumber \\
\VEC H_\LAB{D}' \,  \ISIG[2]^1 \VEC E ~=~&~ 0~, &
\VEC H_\LAB{D}'\, \ISIG[2]^1\ISIG[2]^2 \VEC E
~=~& -\frac{d}2\, \ISIG[2]^1\ISIG[2]^2 \VEC E ~.
\label{eq:eig}
\end{align} 
It follows that the propagator in the transformed basis can be
constructed as
\begin{align} 
\VEC e^{\IMJ \VEC H_\LAB{D}' t}
~=~&~ \HALF \Big( \VEC e^{-\IMJ dt/2}\, \VEC E \,
\begin{aligned}[t]
& -\, \VEC e^{\IMJ dt}\,
\ISIG[2]^2 \VEC E\, \ISIG[2]^2 
\,-\, \ISIG[2]^1 \VEC E\, \ISIG[2]^1
\\ & +\,
\VEC e^{-\IMJ dt/2}\, \ISIG[2]^1\ISIG[2]^2
\VEC E \, \ISIG[2]^1\ISIG[2]^2 \Big)
\end{aligned}
\nonumber \\[-\baselineskip] \\ \nonumber 
~=~&~ \VEC e^{-\IMJ dt/2} \VEC E \,-\, \HALF
\big( 1 + \VEC e^{\IMJ dt} \big)
\ISIG[2]^1 \VEC E \ISIG[2]^1 ~,
\end{align}
where we have used the results
\begin{equation}
\ISIG[2]^1\ISIG[2]^2 \VEC E\, \ISIG[2]^1\ISIG[2]^2 ~=~ \VEC E
\qquad \text{and} \qquad 
\ISIG[2]^2 \VEC E\, \ISIG[2]^2 ~=~ \ISIG[2]^1 \VEC E\, \ISIG[2]^1.
\end{equation}
The propagator is easily transformed back into the original frame
to give 
\begin{align}
\hspace{-0.50em} % much too fat & a little too long...
\VEC e^{-\IMJ \VEC H_\LAB{D}\,t}
~=~&~ \VEC e^{-\IMJ dt/2}\, \VEC E \,
\begin{aligned}[t] -\,
\tfrac14\, & \big( 1 + \VEC e^{\IMJ dt} \big)
\big( \ISIG[2]^{1\,} \VEC E_{\,} \ISIG[2]^1
\,+\, \ISIG[2]^{2\,} \VEC E_{\,} \ISIG[2]^2 \big)
\\ +\,
\tfrac14\, & \big( 1 - \VEC e^{\IMJ dt}
\big) \big( \ISIG[2]^{1\,} \VEC E_{\,} \ISIG[2]^2
\,+\, \ISIG[2]^{2\,} \VEC E_{\,} \ISIG[2]^1 \big)
\end{aligned}
\nonumber \\[-\baselineskip] \\ \nonumber 
~=~&~ \VEC e^{-\IMJ dt/2}\, \VEC E \,-\,
\HALF \big( 1 + \VEC e^{\IMJ dt}
\big) \ISIG[2]^1 \VEC E \ISIG[2]^1 \,+\,
\HALF \big( 1 - \VEC e^{\IMJ dt}
\big) \ISIG[2]^1 \VEC E \ISIG[2]^2 \,,
\label{eq:prop2}
\end{align}
where we have made the same simplifications as before.

\subsection{Observables}

The simple form of the propagator hides a number of interesting
properties, which emerge on forming the observables.   Given
an initial state $\VEC \psi_0$ the state at time $t$ is simply
\begin{equation}
\EMB\psi (t) ~=~ \VEC e^{-\IMJ \VEC H_\LAB{D}\,t}\, \EMB\psi_0 \,.
\end{equation}
Assuming $\|\VEC \psi_0\| = 1$, the spin bivector observable is 
\begin{equation}
\VEC S(t) ~=~ 2\, \EMB\psi(t)\, \VEC J\, \tilde{\EMB\psi}(t) \,,
\end{equation}
which defines separate spin bivectors in spaces 1 and 2.  This object
turns out to have a number of remarkable features.  If the two spins
start out in a separable state with their spin vectors aligned with
the $\LAB z$-axis, i.e.\ $\EMB\psi_0 = \VEC E$, then they do not
evolve since $\VEC H_\LAB{D\,}\VEC E = (-d/2)\VEC E$, collapsing the
propagator to a phase factor.  This result is in accord with the fact
that this orientation is a classically stable equilibrium.

A classically \textit{un}stable equilibrium is obtained when the spins
are aligned antiparallel along the $\LAB z$-axis, e.g.\ in the quantum
state $\EMB\psi_0 = \ISIG[2]^1 \VEC E$.  The time-dependent spinor in
this case is given by
\begin{equation} 
\begin{split}
2\, \VEC e^{-\IMJ \VEC H_\LAB{D\,} t\,}
\ISIG[2]^1 \VEC E
~=~&
\Big( \big( 1 + e^{\,\IMJ dt} \big) \ISIG[2]^1 \,-\,
\big( 1 - e^{\,\IMJ dt} \big) \ISIG[2]^2 \Big) \VEC E
\\ =~&\,
\big( \ISIG[2]^1 - \ISIG[2]^2 \big)
\VEC E \,+\, \big( \ISIG[2]^1 + \ISIG[2]^2 \big)
\big( \cos(dt)\, \VEC E + \sin(dt)\, \VEC J \big) ~,
\end{split} 
\end{equation}
so that the state oscillates between $\ISIG[2]^1\VEC E$ and
$-\ISIG[2]^2\VEC E$ with a period of $2\pi/d$.  This spinor can be
written in the canonical form (obtained by singular value
decomposition \cite{HavelDoran:01})
\begin{equation}
\VEC e^{-\IMJ \VEC H_\LAB{D\,} t\,}
\ISIG[2]^1 \VEC E ~=~ \ISIG[1]^1\, \VEC e^{\IMJ %
\ISIG[2]^1\ISIG[2]^2\,dt/2}\, \VEC E ~,
\end{equation}
from which it is easily seen that the spin bivector is
\begin{equation}
\VEC S(t) ~=~ \cos(dt)\, \big( \ISIG[3]^2 - \ISIG[3]^1 \big) ~.
\end{equation}
Although they remain equal, the magnitudes
of the two spins' bivectors can shrink to zero,
showing that they have become maximally \emph{entangled}.
Thus we see that the state oscillates in and out of entanglement
while swapping the signs of the spin bivectors every half-cycle.

Now suppose that the spins start out with both their vectors parallel
along the $\LAB x$-axis, i.e.\ with $\EMB\psi_0 = \HALF \big( 1 +
\ISIG[2]^1 \big) \big( 1 + \ISIG[2]^2 \big) \VEC E$, which is a saddle
point of the classical energy surface.  Then our propagator gives us
the time-dependent spinor
\begin{equation}
\VEC e^{-\IMJ  \VEC H_\LAB{D\,} t}\, \EMB\psi_0 ~=~
\HALF \Big( \big( 1 + \ISIG[2]^1\ISIG[2]^2 \big) \VEC e^{-\IMJ dt/2}
~+~ \big( \ISIG[2]^1 + \ISIG[2]^2 \big) \VEC e^{\,\IMJ dt} \Big) \VEC E ~,
\end{equation}
which in turn gives rise to the spin bivector observable
\begin{align} 
\VEC S(t) 
~=~&~ \tfrac14\, \big(\ISIG[2]^1 + \ISIG[2]^2 \big)\, \VEC J\,
\VEC e^{\,\IMJ 3dt/2}\, \big(1 + \ISIG[2]^1\ISIG[2]^2 \big) \,-
\nonumber \\ 
& \tfrac14\, \big(1 + \ISIG[2]^1\ISIG[2]^2 \big)\, \VEC J\,
\VEC e^{-\IMJ 3dt/2}\, \big(\ISIG[2]^1 + \ISIG[2]^2 \big)
\\ \nonumber
~=~&~ \cos(3dt/2)\, \big( \ISIG[1]^1 + \ISIG[1]^2 \big) ~.
\end{align}
It follows that the measure of entanglement in this case varies as
the cosine of $\vartheta = 3dt/2$.  Curiously, however, if the spins
start out antiparallel, a similar calculation gives $\vartheta = dt/2$.
Of course the dynamics are the same if the spins
start out (anti)parallel along the $\LAB y$-axis.

\begin{figure}
\begin{center}
\includegraphics[scale=0.67]{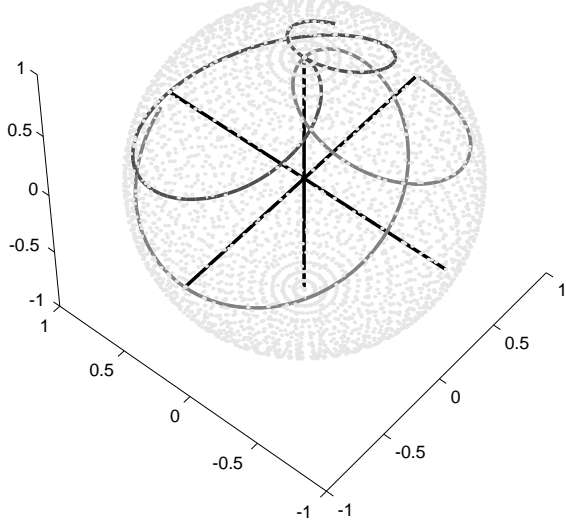}
\end{center}
\caption{Plot of the spin vector trajectories under
the dipolar Hamiltonian starting from an unentangled
state with the first spin (light gray) along the $\LAB
x$-axis and the second (dark gray) along the $\LAB z$-axis.
The length of the vectors stayed near unity,
implying that the state remained largely
unentangled throughout its acrobatics.}
\label{fig:xz}
\end{figure}

Finally, Fig.\ \ref{fig:xz} shows plots of the trajectories of the
spin vectors on (or near) the surface of a unit sphere, starting from
an unentangled state with the first spin (light gray) along the $\LAB
x$-axis and the second (dark gray) along the $\LAB z$-axis.  The
lengths of the spin vectors stayed very nearly at unity, implying that
little entanglement was generated.  The first spin executed a
loop-de-loop up to the $\LAB z$-axis, returned to the $-\LAB x$-axis,
and continued on in the general direction of the $\LAB y$-axis, while
the second swooped down towards the $\LAB y$-axis, returned
symmetrically to the $\LAB z$-axis, and then ended half-way between
the $\LAB x$ \& $\LAB z$-axes.  The complexity of the trajectories
even in such a simple quantum system is impressive!

If we take the derivative of the spin bivector we get
\begin{align}
\partial_{t\,} \VEC S(t) 
~=~&~ 2(-  \VEC H_\LAB{D\,} \EMB\psi \VEC J )  \VEC J  \tilde{\EMB\psi} 
\,+\, 2 \EMB\psi \VEC J (- \VEC H_\LAB{D\,} \EMB\psi \VEC
J)^{\scriptscriptstyle\!\sim\,}  \nonumber \\
~=~&~ 2\big( \VEC H_\LAB{D\,}  \EMB\psi  \tilde{\EMB\psi}
\,-\, \EMB\psi \tilde{\EMB\psi}\, \VEC H_\LAB{D}  \big).
\end{align}
Now $\VEC H_\LAB{D}$ is an MSTA 4-vector, while $\EMB\psi
\tilde{\EMB\psi}$ is the sum of a 4-vector and a scalar.
contains a 4-vector term as well as the scalar term. 
The commutator of two 4-vectors gives rise to terms of grade two,
and it is this that gives rise to the interesting dynamics. 
If the state is separable we can write
\begin{equation}
2\, \bigl\langle \EMB\psi \tilde{\EMB\psi} \bigr\rangle_4 ~=~
\VEC{\iota\hspace{-0.16em}p}^1\VEC{\iota\!q}^2, 
\end{equation} 
where $\VEC{p}$ and $\VEC{q}$ are the single-particle spin vectors.
In this case we have
\begin{equation}
\hspace{0.25em} \partial_{t\,} \VEC S(t) ~=~  \VEC H_\LAB{D} \,
\VEC{\iota\hspace{-0.16em}p}^1\VEC{\iota\!q}^2 - \VEC{\iota\hspace{-0.16em}p}^1\VEC{\iota\!q}^2
\VEC H_\LAB{D} \, .
\end{equation}
For our Hamiltonian the general relation for commutators
of tensor products of three-dimensional bivectors,
\begin{align} 
\big(\VEC A^1\VEC B^2\big) \times \big(\VEC X^1\VEC Y^2\big)
~\equiv~&~ \HALF  \big(\VEC A^1\VEC B^2\big) \big(\VEC X^1\VEC Y^2\big)
- \HALF \big(\VEC X^1\VEC Y^2\big) \big(\VEC A^1\VEC B^2\big)
\nonumber \\
~=~&~ \VEC A \dt \VEC X \,\big(\VEC B \times \VEC Y\big)^2 +
\VEC B \dt \VEC Y \, \big(\VEC A \times \VEC X\big)^1 ~,
\end{align}
enables us to reduce this to
\begin{equation} 
\hspace{-0.71em}
\partial_{t\,} \VEC S(t) 
\,=\, \frac{d}{2}\, (\VEC{p} \wedge \VEC{q})^{1\!}
- \frac{d}{2}\, (\VEC{p} \wedge \VEC{q})^{2\!}
+ \frac{3d}{2}\, \SIG[3] \dt \VEC{q}\, (\SIG[3\!] \wedge \VEC{p})^1 
+ \frac{3d}{2}\, \SIG[3] \dt \VEC{p}\, (\SIG[3\!] \wedge \VEC{q})^2 ,
\!\!\! % much too fat & a little too long...
\end{equation}
and we have recovered exactly the classical equation of motion for two
dipolar-coupled spins.  Of course, separability is not preserved in
the case of quantum spins, due to differences in the higher derivatives. 
This allows the spin vectors to change their lengths (as the
degree of entanglement varies), which cannot happen classically.

\section{Lagrangian Analysis}

Lagrangian methods play a central role in finding equations of motion
for constrained systems such as rigid bodies in classical mechanics.
Via Noether's theorem, they also permit one to identify constants of
the motion such as the total angular momentum, which provide insights
into their long-term dynamics.  Since quantum dynamics typically
exhibit many more symmetries than classical, an extension of
Lagrangian methods to multispin systems is highly desirable.  This
type of analysis is somewhat underutilized in traditional quantum
treatments, as it sits somewhere between classical Lagrangian analysis
and quantum operator techniques.

\subsection{Single-Particle Systems}

As a starting point, consider a single spin-1/2 particle
interacting with an applied magnetic field described by
the bivector $\VEC B$.  The Lagrangian for this is
\begin{equation}
L ~=~ \left\langle \dot{\EMB\psi}\, \ISIG[3]\, \tilde{\EMB\psi} \,-\,
\gamma\VEC B\, \EMB\psi\, \ISIG[3]\, \tilde{\EMB\psi} \right\rangle ~.
\end{equation}
Substituting this into the Euler-Lagrange equations
\begin{equation}
\frac{\FUN d}{\FUN dt} \left( \frac{\partial L}{\partial\dot{\EMB\psi}}
\right) \,=~ \frac{\partial L}{\partial\EMB\psi}
\end{equation}
yields
\begin{equation}
\frac{\FUN d}{\FUN dt} \bigl( \ISIG[3]\,\tilde{\EMB\psi} 
\bigr) ~=~
\bigl(\dot{\EMB\psi}\,\ISIG[3]\bigr)^{\scriptscriptstyle\sim} \,-\,
\gamma\bigl(\ISIG[3]\, \tilde{\EMB\psi}\,\VEC B\bigr) \,-\,
\gamma\bigl(\VEC B\,\EMB\psi\,\ISIG[3] \bigr)^{\scriptscriptstyle\sim} ~.
\end{equation}
This simplifies readily to Schr\"odinger's equation for
the system: 
\begin{equation} 
\dot{\EMB\psi} ~=~ \gamma\VEC B\,\EMB\psi.
\end{equation}
This prototype can be used to build more realistic semi-classical
models of spin, including relativistic effects~\cite{DoLaGuSoCh:96}.
A feature of this Lagrangian, which is typical for spin-$1/2$ systems,
is that it is first order in $\dot{\EMB\psi}$.  For such systems one
frequently finds that $L=0$ for paths satisfying the equations of
motion.

\subsection{Two-Particle Interactions}

Of greater relevance here is the two-particle equation~(\ref{SchE}).
This can be obtained from the Lagrangian 
\begin{equation}
L(\EMB\psi, \dot{\EMB\psi}) ~\equiv~ 2 \left\langle
\dot{\EMB\psi}\, \VEC J\, \tilde{\EMB\psi} \,-\,
\VEC H_\LAB{D}\, \EMB\psi\, \VEC E\, \tilde{\EMB\psi}
\right\rangle ~.
\end{equation}
The first term here can be viewed as the kinetic energy, and the
second term as the potential energy.  The latter term couples entirely
through the 4-vector part of $\EMB\psi\, \VEC E\, \tilde{\EMB\psi}$.
Some insight into the nature of this system can be obtained by
parameterizing the wavefunction as
\begin{equation}
\EMB\psi ~=~ \EMB\upsilon^1 \EMB\zeta^2 (\cos(\theta/2) \EMB E + \sin(\theta/2)
\ISIG[2]^1 \ISIG[2]^2 \EMB J) \VEC e^{\,\alpha \VEC J}.
\end{equation}
Here $\EMB\upsilon$ and $\EMB\zeta$ are single-particle spinors, $\theta$
measures the entanglement and $\alpha$ is an overall phase factor.  In
total this parameterization has 10 degrees of freedom, so 2 must be
redundant.  One of these is in the separate magnitudes of $\EMB\upsilon$
and $\EMB\zeta$, since only their product is involved in $\EMB\psi$.
The second redundant parameter is in the separate single-particle
phases, since under the simultaneous transformation 
\begin{equation}
\EMB\upsilon  \mapsto \EMB\upsilon \, \VEC e^{-\beta \ISIG[3]},
\quad  \EMB\zeta \mapsto \EMB\zeta \, \VEC e^{\beta \ISIG[3]},
\end{equation}
we see that $\EMB\psi$ is unchanged.  Despite this redundancy, the
parameterization is extremely useful,  as becomes clear when we write
the kinetic term as
\begin{equation}
2 \left\langle \dot{\EMB\psi}\, \VEC J\, \tilde{\EMB\psi} \right\rangle 
~=~ \cos(\theta) \bigl\langle \dot{\EMB\zeta} \ISIG[3] \tilde{\EMB\zeta}
\, (\EMB\upsilon \tilde{\EMB\upsilon}) + \dot{\EMB\upsilon} \ISIG[3]
\tilde{\EMB\upsilon} \, (\EMB\zeta \tilde{\EMB\zeta}) \bigr\rangle \,.
\end{equation}
This would reproduce the classical dynamics of two magnetic dipoles,
were it not for the factor of $\cos(\theta)$.  We also see that there
is no derivative term for $\theta$, so the Euler-Lagrange equation for
the entanglement measure produces a simple algebraic equation.

The potential term in the Lagrangian can similarly be written
\begin{align}
- 2 \left\langle \VEC H_\LAB{D}\, \EMB\psi\, \VEC E\, \tilde{\EMB\psi}
\right\rangle  
~=~& - \left\langle \VEC H_\LAB{D}\, 
(\EMB\zeta \ISIG[3] \tilde{\EMB\zeta})^1  
(\EMB\upsilon \ISIG[3] \tilde{\EMB\upsilon})^2 \right\rangle 
\nonumber \\
& +\, \sin(\theta) \left\langle \VEC H_\LAB{D}\, 
\EMB\zeta^1 \EMB\upsilon^2
\bigl( \ISIG[1]^1 \ISIG[2]^2 +  \ISIG[2]^1 \ISIG[1]^2 \bigr)
\tilde{\EMB\zeta}^1 \tilde{\EMB\upsilon}^2 \right\rangle.
\end{align}
Again, it is the entanglement factor that adds the quantum effects.
If we set $\theta=0$ in the Lagrangian then the equations of motion
reduce to precisely those for a pair of classical dipoles.  It is the
presence of $\theta$ in the Lagrangian, which is forced upon us by the
nature of multiparticle Hilbert space, that makes the system truly quantum
mechanical.

\subsection{Symmetries and Noether's theorem} 

Symmetries of the Lagrangian give rise to conserved quantities via
Noether's theorem.  Let $\EMB\psi' = \EMB\psi'(\EMB\psi,\alpha)$ be a
differentiable transformation of the spinor $\EMB\psi$ controlled by a
single scalar $\alpha$, and satisfying $\EMB\psi' (\EMB\psi,0) =
\EMB\psi$.  If we define the transformed Lagrangian as
\begin{equation}
L'(\EMB\psi,\dot{\EMB\psi}) ~\equiv~
L(\EMB\psi',\smash{\dot{\EMB\psi}}') ~,
\end{equation}
we find (using the Euler-Lagrange equations) that
\begin{equation}
\left. \frac{\partial L'}{\partial\alpha} \right|_{\alpha=0}
~=~ \frac{\FUN d}{\FUN dt}\, \left\langle \left.
\frac{\partial\EMB\psi'} {\partial\alpha} \right|_{\alpha=0}\,
\frac{\partial L}{\partial\dot{\EMB\psi}} \right\rangle
~=~ 2 \frac{\FUN d}{\FUN dt}\, \left\langle \EMB\phi\,
\VEC J\, \tilde{\EMB\psi} \right\rangle ~,
\end{equation}
where $\EMB\phi \equiv \partial\EMB\psi'\!/\partial\alpha|_{\alpha=0\,}$.
If $L'$ is independent of $\alpha$ our transformation defines a
symmetry of the Lagrangian, and gives rise to a conjugate conserved
quantity.

As a simple example, take the invariance of $L$ under changes of
phase:
\begin{equation}
\EMB\psi' ~=~ \EMB\psi\,\VEC e^{\,\alpha \VEC J} ~\implies~\;
\EMB\phi ~=~ \EMB\psi\,\VEC J ~.
\end{equation}
It is easily seen that this is a symmetry of the Lagrangian,
and that the conjugate conserved quantity is
$-2\langle\EMB{\psi}\tilde{\EMB\psi}\rangle$,
telling us that the magnitude of $\EMB\psi$ is constant.

Phase changes are something of an exception in that they involve
operation on $\EMB\psi$ from the right.  Operation from the left by a
term of the form $\EMB\exp(\alpha \VEC P)$ will always generate a symmetry
provided $\VEC P$ commutes with the Hamiltonian.  The prime example in
this case is the Hamiltonian $\VEC H_\LAB{D}$ itself.  The symmetry
this generates corresponds is time translation, as defined by the
$\VEC\exp(-\IMJ \VEC H_\LAB{D\,} t)$ and the conjugate conserved
quantity is the total internal energy.  Another important symmetry
generator is provided by $\ISIG[3]^1+\ISIG[3]^2$, which generates
rotation of both spins about the $\LAB z$-axis at equal rates.
(Recall that this axis is defined by the inter-dipole vector.)  This
bivector commutes with the Hamiltonian since
\begin{equation}
\VEC H_\LAB{D}\,\big(\ISIG[3]^1+\ISIG[3]^2\big) 
~=~ -\frac{d}4\,\big(\ISIG[3]^1+\ISIG[3]^2\big) ~=~
\big(\ISIG[3]^1+\ISIG[3]^2\big)\,\VEC H_\LAB{D} ~,
\end{equation}
and the conjugate conserved quantity is 
\begin{equation}
(\ISIG[3]^1+\ISIG[3]^2) \dt (2 \EMB\psi\, \VEC J \, \tilde{\EMB\psi})
= (\ISIG[3]^1+\ISIG[3]^2) \dt \EMB S,
\end{equation}
which gives the total angular momentum about the $\LAB z$-axis.

Yet another symmetry generator is found by considering the operator
for the magnitude of the total angular momentum, namely
\begin{equation} 
\begin{split}
&~
\bigl( \ISIG[1]^1+\ISIG[1]^2 \bigr)^2 \,+\,
\bigl( \ISIG[2]^1+\ISIG[2]^2 \bigr)^2 \,+\,
\bigl( \ISIG[3]^1+\ISIG[3]^2 \bigr)^2
\\ =~&~ 2\,
\bigl( \ISIG[1]^1 \ISIG[1]^2
\,+\,  \ISIG[2]^1 \ISIG[2]^2
\,+\,  \ISIG[3]^1 \ISIG[3]^2
\,-\, 3 \bigr)
~=~ -4\,\big( 1 \,+\, \EMB\Pi \big) ~.
\end{split} 
\end{equation}
The new generator here is $\IMJ \EMB\Pi$, which generates a
continuous version of particle interchange, and is a symmetry of $L$. 
The conjugate conserved quantity is $\langle  \EMB\Pi\, \EMB\psi
\tilde{\EMB\psi} \rangle$.  Rather confusingly, this is \emph{not}
the same as the total magnitude $\langle \EMB S \EMB S \rangle$,
which is \textit{not} conserved in the quantum case (though it is
conserved classically where the spin vectors have fixed length).

Together with phase invariance, we have now found a total of four
physically relevant constants of motion for the two-spin dipolar
Hamiltonian $\VEC H_\LAB{D\,}$.  Are there any more?  Clearly any
linear combination of constants of motion is again a constant of
motion, so this can only be answered in the sense of finding a
complete basis for the subspace spanned by the constants of motion in
the $16$-dimensional space of all observables.  The dimension of this
subspace can be determined by considering the eigenstructure of $\VEC
H_\LAB{D}$ again (Eq.\ \ref{eq:eig}).  Clearly the observable which
measures the amount of a state in any one eigenspinor is a constant of
the motion, and more generally, each eigenvalue gives rise to
$\delta^2$ constants of the motion where $\delta$ is its degeneracy.
Since $\VEC H_\LAB{D}$ has one two-fold degenerate eigenvalue,
the constants of motion must span a subspace of dimension
$1 + 1 + 2^2 = 6$, and we are therefore just two short! 
It is quite easy to see, however, that the so-called
(in NMR) \emph{double quantum} coherences
\begin{equation}
\ISIG[1]^1\ISIG[1]^2 \,-\, \ISIG[2]^1\ISIG[2]^2
\quad\text{and}\quad
\ISIG[1]^1\ISIG[2]^2 \,+\, \ISIG[2]^1\ISIG[1]^2
\end{equation}
commute with $\VEC H_\LAB{D}$ and
generate two new independent symmetries.
The physical interpretation of these symmetries is far from
simple as they again involve the geometry of the 4-vector
$\langle \EMB\psi\, \tilde{\EMB\psi} \rangle_4$. 
Because the product of any two constants of motion is again
a constant of motion, the constants of motion constitute
a \emph{subalgebra} (of the even subalgebra) of the MSTA.
This makes it possible to form large numbers of new constants
in order to find those with the simplest physical or geometric
interpretations.

\section{The Density Operator}

For a single-particle normalized pure state
the quantum density operator is
\begin{equation}
\hat{\rho} ~=~ \KET{\psi} \BRA{ \psi} ~=~ \HALF\, \big(\hat{I} +
{\textstyle\sum}_{k=1\,}^3 \, p_k\, \hat{\sigma}_k\big) \,,
\end{equation}
where $\hat{I}$ is the identity operator
on a two-dimensional Hilbert space, and 
\begin{equation}
p_k ~=~ \BRA{\psi} \,\hat{\sigma}_k\, \KET{\psi} \,.
\end{equation}
The MSTA equivalent of this is simply
\begin{equation}
\EMB\psi\, \VEC Z_{+\,} \tilde{\EMB\psi} ~=~ \HALF (1 + \VEC{p}) ~,
\end{equation}
where $\VEC Z_\pm \equiv \HALF(1 \pm \SIG[3])$ are idempotents
and $\VEC{p} = \VEC{\psi} \,\SIG[3]\, \VEC{\tilde{\psi}}$ is
known as the \emph{Bloch} or spin \emph{polarization} vector.

The most straightforward extension of this to the multiparticle
case replaces the idempotent by the product of those for all the $n$
distinguishable spins: $\VEC Z_+ \equiv \VEC Z_+^1\cdots\VEC Z_+^n$.
This has the added benefit of enabling one to absorb all
the $\SIG[3]^a$ vectors from $\VEC E$ into $\VEC Z_{+\,}$,
thereby converting $\VEC E$ into a pseudoscalar correlator,
\begin{equation}
\VEC C ~=~ \HALF( 1 - \EMB{\iota}^1\EMB{\iota}^2 )
\cdots \HALF( 1 - \EMB{\iota}^1 \EMB\iota^n ) ~.
\end{equation}
This commutes with the entire even MSTA subalgebra, and allows one to
identify all the pseudoscalars $\EMB\iota^a$, wherever they occur,
with a single global imaginary as in conventional quantum mechancs.
This faithfully reproduces all standard results, but the idempotent
$\VEC Z_+^1\cdots\VEC Z_+^n$ does not live in the even subalgebra, and
hence mixes up the grades of entities which otherwise would have a
simpler geometric meaning.  In the following we propose for the first
time a formulation of the density operator within the even subalgebra
of the MSTA.

The key is to observe that both $\VEC Z_+^1\cdots\VEC Z_+^n$
and $\VEC E + \VEC J$ contain all $2^n$ possible
products of the $\SIG[3]^a$ ($a = 1,\ldots,n$).
It follows that the even MSTA density operator
\begin{equation}
\RHO_* ~\equiv~ 2^{n-1}\, \EMB\psi\,
\big( \VEC E \,+\, \VEC J \big)\, \tilde{\EMB\psi}
\end{equation}
includes the same complete set of commuting observables
as does the usual definition (modulo pseudoscalar factors).
The normalization of $2^{n-1}$ is chosen so that
the scalar part $\langle \RHO_* \rangle = 1$.
This is more natural when using geometric algebra
than the usual factor of $1/2$, which ensures that
the trace of the identity state $\VEC I/2^n$ is unity.
The density operator of a mixed state is simply the
statistical average of these observables as usual:
$\RHO_* \equiv \OL{\EMB\psi(\VEC E \,+\, \VEC J)
\tilde{\EMB\psi}}$ \cite{HaCoSoTs:00,HavelDoran:01}.

There is a problem with the even MSTA version, however, which is
that $\EMB\psi\, \EMB J \tilde{\EMB\psi}$ is \textit{anti}-Hermitian
(under reversion), whereas the usual density operator is Hermitian.
In addition, up to now we have only applied the propagator to spinors,
where the complex structure is given by right-multiplication with $\EMB J$.
Such a representation is not appropriate for density operators,
because to apply it we would have to decompose our density
operator into a sum over an ensemble of known pure states,
apply the propagator to each state's spinor, and then
rebuild the density operator from the transformed ensemble.
This is clearly undesirable, so we should re-think the role
of the imaginary in the propagator for the density operator.

To do this, we must first understand the role of $i$ in the
propagator, where it appears multiplying the (Hermitian) Hamiltonian
$\hat{H}$.  The terms in $i \hat{H}$ involving odd numbers of
particles have an immediate counterpart in the MSTA as products of odd
numbers of bivectors.  These exponentiate straightforwardly in the
MSTA, and here the role of the imaginary as a pseudoscalar is clear.
It is the terms in $i \hat{H}$ involving even numbers of particles
which are the problem.  When these are converted to products of even
numbers of bivectors, a single factor of $i$ is left over.  To see its
effect we return to a pure state and consider
\begin{equation}
(\IMJ \EMB \psi) \tilde{\EMB\psi} ~=~ \EMB\psi\, \EMB J \tilde{\EMB\psi}
\quad \mbox{and} \quad
(\IMJ \EMB \psi) \EMB J \tilde{\EMB\psi} ~=~ - \EMB\psi\, \tilde{\EMB\psi} \,.
\end{equation}
Thus we see that, applied to density operators (or any other
observables), the imaginary unit interchanges even and odd terms (in
their number of bivectors), in addition to squaring to $-1$.

It follows that the left-over factor of $i$ in the even terms
converts them to odd terms, thereby ensuring that the Hamiltonian
generates a compact group while at the same time \emph{labeling}
these terms to keep them separate from the original odd terms.
This is quite different from other occurrences of the imaginary!
A simple way to represent this in the even MSTA is to introduce a ``formal''
imaginary unit $j$, and to redefine the even MSTA density operator as
\begin{equation}
\RHO ~\equiv~ \RHO_+ \,-\, j\, \RHO_-
\end{equation}
where $\RHO_+ \equiv {\langle\RHO_*\rangle}_0
+ {\langle\RHO_*\rangle}_4 + \cdots$ and
$\RHO_- \equiv {\langle\RHO_*\rangle}_2
+ {\langle\RHO_*\rangle}_8 + \cdots$.
Then, provided any residual factors of the even MSTA
operator $\IMJ$ are replaced by $j$, everything works simply.
In writing $\RHO$ in this way we have also
recovered a more standard, Hermitian representation,
which permits the mean expectation values of any
observable $\EMB{o} = \EMB{o}_+ + j\EMB{o}_-$
to be computed by essentially the usual formula:
\begin{equation}
\langle\, \EMB o\, \RHO \,\rangle_0 ~=~ \langle\,
(\EMB{o}_+ \,+\, j\,\EMB{o}_-)(\RHO_+ \,-\, j\,\RHO_-) \,\rangle_0 ~=~
\langle\, \EMB{o}_{+\,}\RHO_+ \,+\, \EMB{o}_{-\,}\RHO_- \,\rangle_0 ~.
\end{equation}
The geometric interpretation of this new imaginary operator
$j$ will be further considered on a subsequent occasion.

\subsection{An Example of Information Dynamics}

We are now ready to see how things look to one spin when no
information about the other(s) is available.  Clearly this depends on
the (unknown!) configuration of the other spin(s), and we expect that
the ``generic'' case will result in a very complicated trajectory,
which may also be \emph{incoherent} in the sense that the length of
the spin vector is not preserved.  Somewhat surprisingly, the
situation is simplified substantially by assuming that the spin of
interest has an environment consisting of a great many others, whose
dynamics are so complex that their spin vectors can be treated as
completely random.  Under these circumstances, neither the exact
starting state of the environment, nor any subsequent change in it,
will change the way our chosen spin (henceforth numbered $1$) sees it.
As a result of this assumption and the linearity of quantum dynamics,
we need only figure out how its basis states $\ISIG[1]^1, \ISIG[2]^1,
\ISIG[3]^1$ evolve in order to determine its evolution in general.

To see how the  $\ISIG[k]^1$ evolve, we return to the propagator of
equation~(\ref{Prop1}), which we now write in the form
\begin{equation} 
\EMB\exp\big( \!-\! j\, \VEC H_\LAB{D\,} t \big) ~=~
\EMB\exp\big( \!-\! j\, \ISIG[3]^1 \ISIG[3]^2\, 3dt/4\big) \,
\EMB\exp\big( \!-\! j\, \EMB\Pi\, dt/2 \big) \,
\EMB\exp\big(j_{\,}dt/4\big) \,. 
\end{equation}
The phase term has no effect, and for the interchange term we find
that
\begin{align}
\VEC e^{ -j\, \EMB\Pi\, dt/2 }\, \ISIG[k]^{1}\,
\VEC e^{ \, j\, \EMB\Pi\, dt/2 } 
~=~&~ \cos^2(dt/2)\, \ISIG[k]^1 + \sin^2(dt/2)\, \ISIG[k]^2 
\nonumber \\
&\qquad -\, j\, \sin(dt) \,\EMB \Pi \times  \ISIG[k]^1 \,.
\label{Swap}
\end{align}
We then need to apply the term in $\EMB\exp( j\, \ISIG[3]^1 \ISIG[3]^2\,
3dt/4)$.  This commutes with $\EMB \Pi$, so we only need to transform
the separate bivectors.  For the case of $k=3$ this final term has no
effect, and we have
\begin{align} 
\VEC e^{-j_{\,} \VEC H_\LAB{D\,} t}\,
\ISIG[3]^1\,
\VEC e^{\,j_{\,} \VEC H_\LAB{D\,} t}
~=~&~
\HALF \big( 1 + \cos(dt) \big) \ISIG[3]^1 \,+\,
\HALF \big( 1 - \cos(dt) \big) \ISIG[3]^2 
\nonumber \\
& \qquad -\, j\, \sin(dt)\, \HALF \big( \ISIG[1]^1\ISIG[2]^2
 - \ISIG[2]^1\ISIG[1]^2 \big) .
\end{align}
For $k=1,2$ we need the result
\begin{equation}
\VEC e^{ - j_{\,} \ISIG[3]^1 \ISIG[3]^2\, \phi/2\, } \ISIG[2]^1 \,
\VEC e^{ \, j_{\,} \ISIG[3]^1 \ISIG[3]^2\, \phi/2 } ~=~ \cos(\phi)\,
\ISIG[2]^1 - j\, \sin(\phi)\, \ISIG[1]^1 \ISIG[3]^2 \,.
\end{equation}
Combining this with the preceding we see that
\begin{align} 
\VEC e^{-j_{\,} \VEC H_\LAB{D\,} t}\,
\ISIG[2]^1\, \VEC e^{\,j_{\,} \VEC H_\LAB{D\,} t}
~=~&
\big(\! \cos(dt){\scriptstyle\,} \ISIG[2]^1 -
\sin(dt){\scriptstyle\,} j\,\ISIG[1]^1\ISIG[3]^2 \big) \cos(dt/2)
\nonumber \\ 
& - \big(\! \cos(dt){\scriptstyle\,} j\,\ISIG[3]^1\ISIG[1]^2 +
\sin(dt){\scriptstyle\,} \ISIG[2]^2 \big) \sin(dt/2) \, ,
\end{align}
with a similar result holding for $\ISIG[1]^1$.

As a simple model, suppose that our initial state is described by a
known state of particle~1, encoded in the density operator $\RHO(0) =
(1 + \VEC p(0) )/2$, and with the state of the second spin taken to be
totally random.  To evolve the density matrix of particle~1 we write
$\EMB\iota\!\VEC p(0) = \sum_{k\,} p_k \ISIG[k]^1$ and evolve each
term as above.  We then project $\RHO(t)$ back into the first
particle space (by throwing out any terms involving other spins)
and reform the polarization vector.   We then find that:
\begin{equation}
\VEC p(t) ~=~ \cos(dt)\cos(dt/2)\, \big(p_{1\,} \SIG[1] ~+~
p_{2\,} \SIG[2] \big) \,+\, \cos^2(dt/2)\, p_{3\,} \SIG[3] ~.
\end{equation}
Of course, the same result is obtained if everything is rotated
to some other orientation, for example if the other spin is along
$\LAB x$ and $\VEC p$ is replaced by the rotated spin vector.

\begin{figure}
\begin{center}
\includegraphics[height=2in,width=4in]{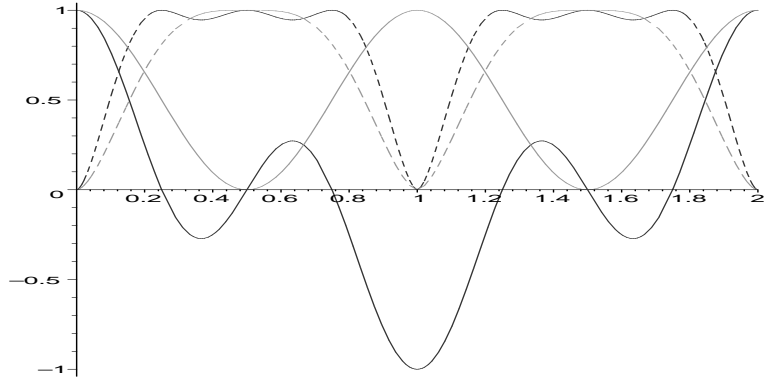}
\end{center}
\caption{Dipolar ``tides'' (in units of $1/d$) exerted
on a spin's polarization vector by its interactions
with a random ensemble of other spins all located in a
circular ``orbit'' about its equator (light gray line),
or at a fixed distance above its poles (dark gray line),
along with the corresponding von Neumann entropies
(dashed lines).} \label{fig:tides}
\end{figure}

The von Neumann entropy of the spin, which measures how much
information about its state has been lost to the environment,
is given (in bits) by
\begin{align}
S_\LAB{vN} ~=~ 
& -\HALF (1 + \|\VEC p(t)\|) \log_2\big( (1 + \|\VEC p(t)\| )/2 \big)
\nonumber \\ 
& - \HALF(1 - \|\VEC p(t)\|) \log_2\big( (1 - \|\VEC p(t)\| )/2 \big) ~.
\end{align}
The signed lengths of the spin polarization
vectors and the corresponding von Neumann entropies
are plotted in Fig.\ \ref{fig:tides} for a second spin
in a random state and displaced from the first in directions
parallel and perpendicular to its polarization vector.

\subsection{Towards Quantum Complexity and Decoherence}

Our applications of geometric algebra to the problem of two
dipole-coupled spins has illustrated how quantum interactions allow
qubits to exchange information (states), in the process passing
through an entangled state in which no information on either is
directly accessible.  We have also seen how information on the state
of a qubit can be lost through its interactions with an environment,
although the Poncaire recurrence times for the simple one-qubit
environments we have considered are short enough to allow us to also
see the underlying periodic behavior.  This constitutes the simplest
case of an old and venerable problem in solid-state NMR spectroscopy,
which is to predict the spectral line-shape (decay envelope) of a
\emph{macroscopic} system of dipole-coupled spins such as calcium
fluoride \cite{Abragam:61,Cowan:97,SodicWaugh:95}.  Here an exact
treatment is out of the question, since the analytic evaluation of
even the three-spin propagator is a reasonably challenging (though
solvable) problem.  Nevertheless, we believe the insights provided by
geometric algebra, particularly into the constants of the motion,
offer the hope of further progress.

More generally, quantum mechanical explanations of apparently
irreversible processes are becoming increasingly important in many
fields \cite{Weiss:99}, and further have the potential to finally
clarify in just what sense classical mechanics can be regarded as a
limiting case of quantum mechanics \cite{GiuliniEtAl:96}.  These
problems typically involve the spatial degrees of freedom, and hence
are truly infinite dimensional.  Most of what is known about
\emph{decoherence} in such systems therefore comes from the analysis
of simple and highly tractable environmental models such as a bath of
spins or harmonic oscillators.  Because of its potential to provide
global insights into the full spacetime structure of such models, we
expect that the MSTA will also come to play an enabling role in our
understanding of the quantum mechanical mechanisms operative in
decoherence.

\bigskip\section*{Acknowledgements}
TFH thanks Prof.\ David Cory of MIT for useful discussions on NMR,
and ARO grant DAAG55-97-1-0342 \& DARPA grant MDA972-01-1-0003
for financial support.  CJLD is supported by the EPSRC.

\end{document}